\documentstyle[emulateapj,epsf]{article}

\begin{document}

\title{ The Role of Clustering of Sub-Clumps
in Bright Elliptical Galaxy Formation
from a Low-Spin Seed Galaxy }
\author{Daisuke Kawata}
\affil{Astronomical Institute, Tohoku University, Sendai,
980-8578, Japan}
\authoremail{kawata@astr.tohoku.ac.jp}

\begin{abstract}

We reveal the role of clustering of sub-clumps, which is
expected in the cold dark matter (CDM) universe,
in forming a bright elliptical galaxy (BEG) from a low-spin seed galaxy.
This can be done by comparing the evolution of a low-spin seed galaxy
including small-scale density fluctuations expected in the CDM universe
(Model 1) with that of a completely uniform one (Model 2),
using numerical experiments.
We show that Model 2 cannot reproduce the properties of BEGs
and forms a disk which is too compact and too bright
due to the conservation of the initial-small angular momentum.
In Model 1 clustering of the sub-clumps caused by initial 
small-scale density fluctuations leads to
angular momentum transfer from the baryon component to the dark matter
and consequently a nearly spherical system supported
by random motions is formed. Moreover the collisionless property
of the stars formed in the sub-clumps
prevents the dissipative contraction of the system,
leading to a large measured half-light radius. 
As a result, the end-product is quite well reproduces
the observed properties of BEGs,
such as the de Vaucouleurs light-profile, typical color and metallicity
gradients, the large half-light radius,
the small ratio of the rotational velocity
to the velocity dispersion ($V/\sigma$).
We conclude that the clustering of sub-clumps, i.e.,
the hierarchical clustering, plays a crucial role
in the formation of BEGs from a low-spin seed galaxy.
\end{abstract}


\section{Introduction}

 Observed bright elliptical galaxies (BEGs) have the common interesting
properties, which are expected to provide strong constraints on
their formation scenario.
For example, Davies et al.\ (1983) found that the ratios of the rotation
velocity to the velocity dispersion, $V/\sigma$, of BEGs are much smaller than
those of faint ellipticals.
In photometric properties, the light profile of BEGs is
well described by the de Vaucouleurs law,
whereas faint ellipticals usually have an exponential light profile
(e.g., Caon, Capaccioli, \& D'Onofrio 1993). 

 On the other hand, Kawata (1999) showed that in the CDM universe,
i.e., the hierarchical clustering scenario,
elliptical galaxies can be formed from a low-spin overdense region,
using numerical simulation
which includes almost all the important physical processes
in galaxy formation self-consistently. He adopted a semi-cosmological
model proposed by Katz \& Gunn (1991).
In this model, the process of galaxy formation is mimicked by
a collapse of a top-hat over-dense sphere (a seed galaxy)
which initially follows a Hubble flow expansion and has
a solid-body rotation as an effect of the external tidal field.
The amount of the initial solid body rotation is specified by a
dimensionless spin parameter, $\lambda$ (Peebles 1971).
In addition, small-scale density fluctuations expected
in the CDM universe are superposed on this sphere
which consists of dark and baryonic matter.
Using this model, Katz \& Gunn (1991), Katz (1992),
and Steinmetz \& M\"uller (1994, 1995) studied only the case of
$\lambda=0.08$ and seemed to succeed in forming a system
which has similar properties to an observed disk galaxy
in several respects (see also Koda, Sofue, \& Wada 2000a; 2000b, who presented,
in the same way, that disk-like systems are
formed for a range of $\lambda=0.10-0.04$).
However, the spin parameter of the virialized dark matter
falls in a range 0.02--0.11 with a median value of 0.05,
according to the results of N-body simulations
(e.g., Barnes \& Efstathiou 1987; Warren et al.\ 1992)
 and analytical works (e.g.,
Steinmetz \& Bartelmann 1995; Eisenstein \& Loeb 1995),
in which $\lambda=0.08$ is near a high-end value.
Kawata (1999) studied the evolution of a seed galaxy which
has a slow rotation corresponding to $\lambda=0.02$
and showed that the end-product reproduces the observed
properties of elliptical galaxies very well.
Cosmologically, a small spin-parameter is preferred
in a galactic halo collapsing at a high redshift
(Heavens \& Peacock 1988). Moreover, elliptical galaxies
are considered to be a system whose stellar component is old
and rotates slowly. Thus, it is a quite natural consideration
that most elliptical galaxies were formed in a halo
collapsing at high redshift and rotating slowly.

\begin{figure*}
\epsscale{.9}
\plotone{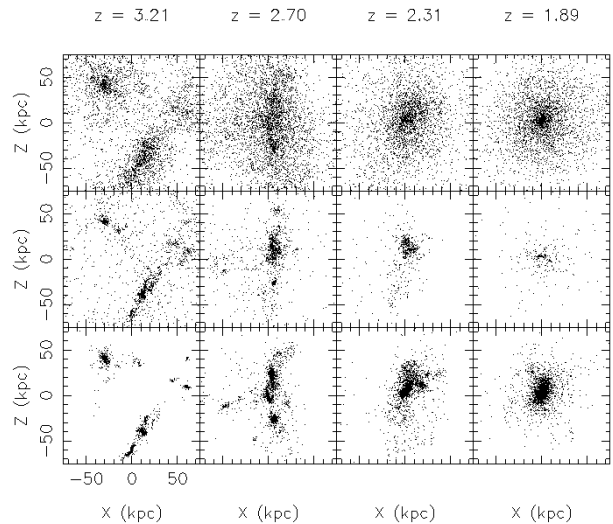}
\figcaption[f1.eps]
{ Time evolution of the system in Model 1. The upper, middle, and lower
panels show the distributions of the dark matter, gas, and stars
respectively.  Each panel measures 150 kpc across and shows the $x$-$z$
projection of the particles, where we set the $z$-axis to be the initial
rotation axis.
\label{anim1}}
\end{figure*}

Kawata's study was based on the hierarchical clustering scenario,
because in his simulation the galaxy forms through mergers of sub-clumps
caused by initial small-scale density fluctuations
(see Fig.\ 1 of Kawata 1999). The question we have to ask here is
how the clustering of sub-clumps acts on the formation of elliptical
galaxies from a low-spin seed galaxy.
In order to understand further the role of the clustering,
we compare the evolution of a low-spin seed galaxy
including small-scale density perturbations with that of a uniform one,
as the opposite case. Then, we focus on massive seed galaxies, and
examine whether both types of seed galaxy can reproduce the observed
properties of BEGs quantitatively, because
the properties of BEGs, as mentioned above,
are expected to be able to provide constraints on their formation
scenario.
If the uniform low-spin seed galaxy also succeeds in forming a system
similar to BEGs, small-scale density fluctuations superposed
on the initial slow-rotating sphere will not be indispensable to the
model. The present study will thus clarify the role of the clustering of
the sub-clumps induced by small-scale density fluctuations
in galaxy formation from a massive low-spin seed galaxy.

 The plan for the remainder of the paper is as follows.
In Section 2, the numerical method and the model are described.
Section 3 presents results of numerical simulations.
Conclusion is described in Section 4.

\begin{figure*}
\epsscale{.9}
\plotone{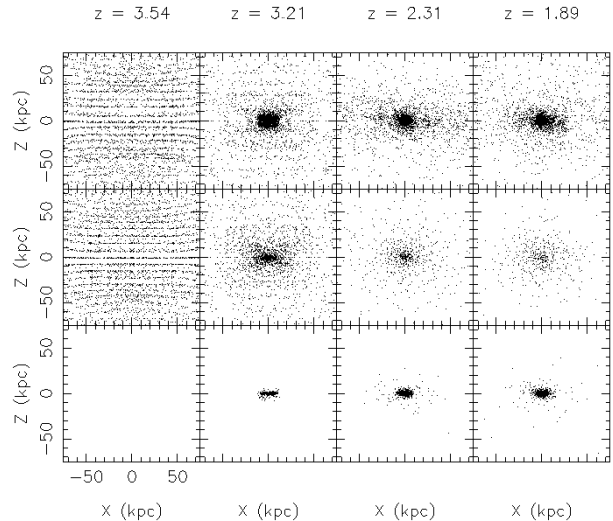}
\figcaption[f2.eps]
{ The same as Fig.\ 1 but for Model 2.
\label{anim2}}
\end{figure*} 

\section{Methods}

\subsection{The Code}

Details of our code are described in Kawata (1999).
It is essentially based on the TreeSPH
(Hernquist \& Katz 1989; Katz, Weinberg, \& Hernquist 1996),
which combines the tree algorithm (Barnes \& Hut 1986)
for the computation of the gravitational forces with the smoothed
particle hydrodynamics (SPH; Lucy 1977; Gingold \& Monaghan 1977)
approach to numerical hydrodynamics.
The dynamics of the dark matter
and stars is calculated by the N-body scheme, and the
gas component is modeled using the SPH.
It is fully Lagrangian, three-dimensional, and highly adaptive in space
and time owing to individual smoothing lengths and
individual time steps.
Moreover, it self-consistently
includes almost all the important physical processes
in galaxy formation, such as self-gravity, hydrodynamics,
radiative cooling, star formation, supernova feedback, and
metal enrichment.

The radiative cooling which depends on the metallicity
(Theis, Burkert, \& Hensler 1992)
is taken into account. The cooling rate of a gas with the solar metallicity
is larger than that for a gas of the primordial composition
by more than an order of magnitude.
Thus, the cooling by metals should not be ignored
in numerical simulations of galaxy formation
(K\"aellander \& Hultman 1998; Kay et al.\ 2000).

The star formation occurs where the gas density is greater than
a critical density, $\rho_{\rm crit}=7\times10^{-26} {\rm g\ cm^{-3}}$,
and the gas velocity field is convergent.
The star formation rate (SFR) is given by
\begin{equation}
 \frac{d \rho_*}{dt} = -\frac{d \rho_{g}}{dt}
 = \frac{c_* \rho_g}{t_g},
\label{sfreq}
\end{equation}
 where $c_*=0.5$ is a dimensionless SFR parameter and
$t_g = \sqrt{3 \pi/16 G \rho_g}$ is the dynamical time,
which is longer than the cooling time
in the region eligible to form stars.
This formula corresponds to a Schmidt law in which SFR
is proportional to $\rho_{g}^{1.5}$.
We apply a Salpeter IMF with the lower mass limit of $0.1 M_\odot$
and the upper mass limit of $60 M_\odot$.
After stars have formed, the massive stars ($>8M_{\odot}$) die immediately,
accompanied by supernova explosions which inject
thermal energy and metals to the surrounding gas.
We adopt the same stellar nucleosynthesis yields 
as Kodama \& Arimoto (1997).
Type Ia supernovae are not taken into account in this paper.

\subsection{The Model}

 We use the models which are almost the same as Kawata (1999).
We consider an isolated sphere as a seed galaxy and
compare the two models whose initial conditions are different.
The only one difference between the two models is whether
small-scale density fluctuations following a CDM power spectrum
are superimposed on the initial isolated sphere (hereafter Model 1)
or not (hereafter Model 2).
In Model 1 the small-scale CDM density fluctuations are generated
by Bertschinger's software COSMICS (Bertschinger 1995).
To incorporate the effects of fluctuations with longer wavelengths,
the density of the initial sphere has been enhanced and a rigid rotation
corresponding to a spin parameter, $\lambda$, has been added.
The initial condition is determined
by four parameters $\lambda$, $M_{\rm tot}$,
$\sigma_{\rm 8,in}$, and $z_{\rm c}$:
the spin parameter is defined by
\begin{equation}
 \lambda \equiv \frac{J|E|^{1/2}}{G M_{\rm tot}^{5/2}},
\end{equation}
where $J$ is the total angular momentum of the system,
$E$ is the total energy,
and $M_{\rm tot}$ is the total mass of this sphere, which is
initially composed of dark matter and gas;
$\sigma_{\rm 8,in}$ is defined only in Model 1 and specifies
an rms mass fluctuation in a sphere of radius
$8\ h^{-1}$ Mpc, which normalizes the amplitude of the CDM power spectrum;
$z_c$ is the expected collapse redshift.
 If the top-hat density perturbation has an amplitude of
$\delta_i$ at an initial redshift, $z_i$, we obtain
$z_{\rm c} = 0.36 \delta_i (1+z_i)-1$ approximately
(e.g., Padmanabhan 1993).
Thus, when $z_{\rm c}$ is given, $\delta_i$ at $z_i$ is determined.
Since we focus on the evolution of a massive low-spin seed galaxy,
we set these parameters,
$\lambda=0.02$, $M_{\rm tot}=4\times10^{12} M_{\odot}$,
and $z_c=2.07$, which corresponds to an over-density of $2.5 \sigma$
for a CDM power spectrum with $\sigma_8=0.5$ (White \& Frenk 1991).
In Model 1 we employ $\sigma_{\rm 8, in}=0.5$.
In Model 2 the initial sphere does not have any small-scale density
fluctuations, i.e., the sphere has a completely uniform density.
We assume a flat universe ($\Omega=1$) with a
baryon fraction of $\Omega_b=0.1$ and a Hubble constant
of $H_0 = 50\ {\rm km\ s^{-1}\ Mpc^{-1}}$.

\begin{center}
\epsscale{.4}
\plotone{f3.eps}
\figcaption[f3.eps]
{ Time variations of the star formation rate (upper panel),
angular momentum (middle panel),
 and number of stellar clumps (bottom panel) in Model 1.
In the middle panel, the thick solid line represents
the evolution of the angular momentum of the baryon (gas and stars),
and the solid and dotted lines show
the time variation of the angular momentum of the gas and stars
respectively. The angular momentum is not a specific one but
the total value for each component.
\label{h1}}
\end{center}

 We carry out these two simulations using 9171 particles for
gas and dark matter respectively. The masses of individual
gas, stellar, and dark matter
particles are $4.36\times10^7$, $4.36\times10^7$,
and $3.93\times10^8 M_\odot$ respectively,
and softening lengths of gas, stellar, and dark matter particles
are $2.06$, $2.06$, and $4.28\ {\rm kpc}$
respectively.

\section{Results}

\subsection{Time Evolution}

 We simulated the evolution of the two models from $z_i=40$ to $z=0$.
Figs.\ \ref{anim1} and \ref{anim2}
show the evolution of the system in Model 1 and Model 2 respectively.
These panels show the projection of the particles onto the $x$--$z$ plane,
where we take the $z$-axis to be the initial rotational axis.
The evolution in Model 1 resembles Fig.\ 1 of Kawata (1999),
although he simulated the evolution of a lower mass system
($M_{\rm tot}=8\times10^{11} M_\odot$) than that of this paper.
In Fig.\ 1 the system has already
turned around at $z=3.21$. We can observe the sub-clumps caused by
initial small-scale density fluctuations. At $z=2.31$, these
clumps merge at the center of the system.
The whole system has collapsed and
settled into a spherical system at $z=1.89$.
After that, the system evolves little morphologically.
The evolution of Model 2 is quite different (Fig.\ \ref{anim2}).
The gas is accreted to the center very smoothly,
which induces star formation. Although
the initial angular momentum is very small,
a disk-like object is formed by
contraction along the initial rotation axis.
Model 2 also evolves little morphologically from $z=1.89$ to $z=0$.

\begin{center}
\epsscale{.4} 
\plotone{f4.eps} 
\figcaption[f4.eps]
{ Time variations of the star formation rate (upper panel)
 and angular momentum (lower panel)
 in Model 2.
 In the lower panel, the thick solid line represents
the evolution of the angular momentum of the baryon (gas and stars),
and the solid and dotted lines show
the time variation of the angular momentum of the gas and stars
respectively. The angular momentum is not a specific one but
the total value for each component.
\label{h2}}
\end{center}

\begin{figure*}
\epsscale{.9}
\plotone{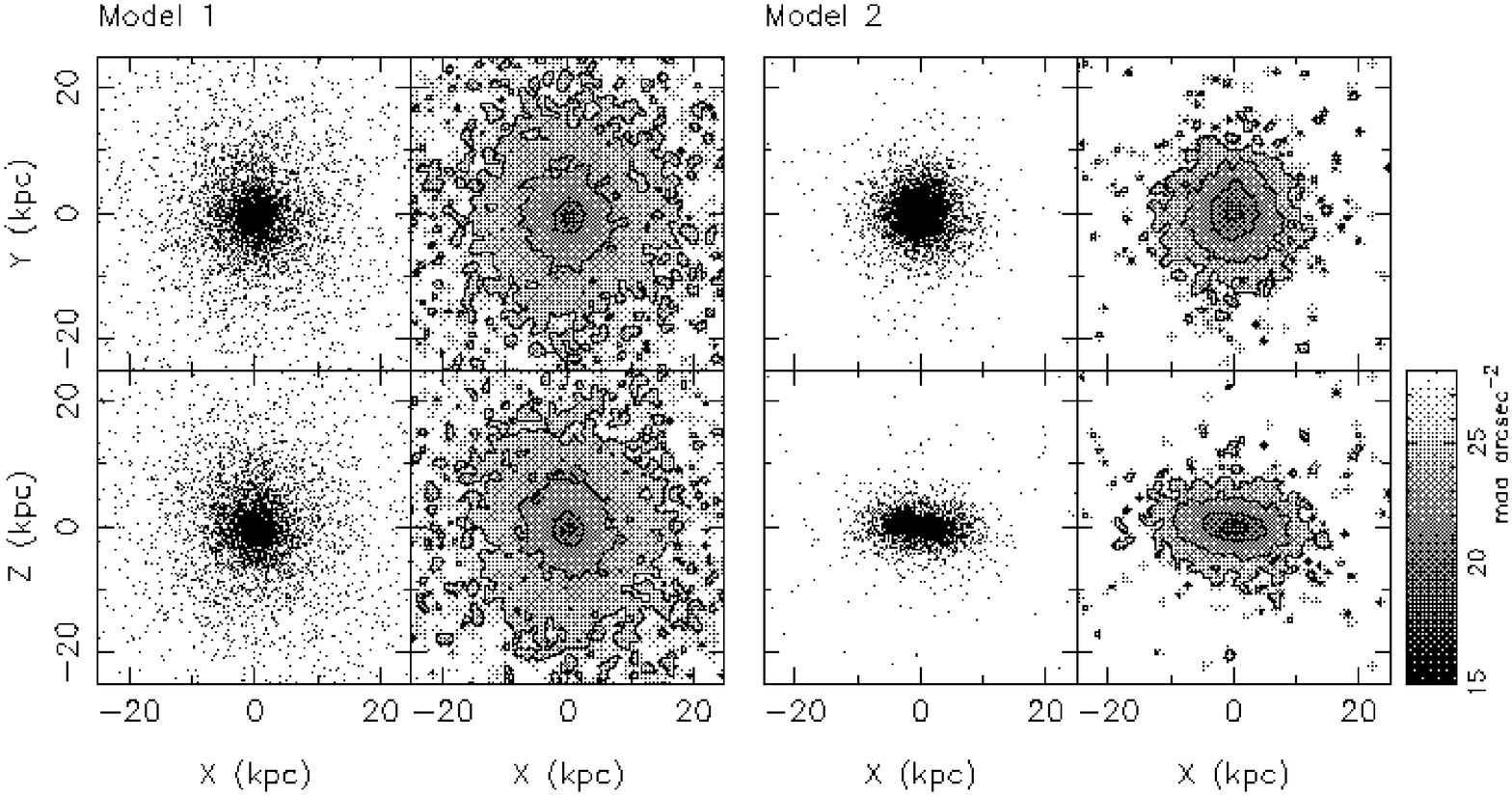}
\figcaption[f5.eps]
{
 Final distribution of stellar particles (left) and $B$ band image (right)
 for Model 1 and Model 2. The upper (lower) panels show the $x$-$y$
 ($x$-$z$) projections, where we take the $z$-axis to be
 the initial rotational axis. Each panel measures 50 kpc across.
 The contours correspond to the isophotes of 25, 23, 21, and 19
 mag arcsec$^{-2}$ respectively.
\label{last}}
\end{figure*}

 Figs.\ \ref{h1} and \ref{h2} show the time variation
of the star formation rate, the angular momentum of the baryon component,
and the number of the stellar clumps in Model 1 and Model 2 respectively. 
Here, a clump means a bound object which is identified by applying
to the stellar particles the friend-of-friend (FOF) algorithm with
a linking length of 0.2 times the particle separation
required to attain the mean density of the universe
at each redshift and with a threshold particle number of 10.
Since Model 2 initially has no internal
small-scale density fluctuation causing formation of sub-clumps
and its subsequent structure is also very smooth (Fig.\ \ref{anim2}),
we do not plot the history of the number of the clumps in Fig.\ \ref{h2}.

\begin{figure*}
\epsscale{.9}
\plotone{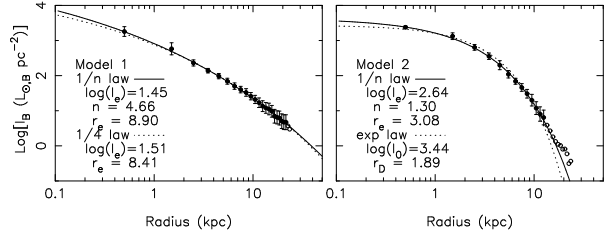}
\figcaption[f6.eps]
{
The $B$ band surface brightness profiles in Model 1 (left) and Model 2 (right).
The solid lines denote the $r^{1/n}$ law best-fit
for the data plotted as the solid symbols.
In the left (right) panel, the dotted line denotes the $r^{1/4}$
(exponential) law best-fit for the solid symbols.
The fitting parameters are shown in the lower left corner of each panel.
The error bars are shown only for the solid symbols
and correspond to the standard deviation.
\label{sfb}}
\end{figure*}

 In Model 1 star formation begins in the sub-clumps before the whole system
collapses. The major star formation occurs
due to the collapse of the whole system and continues for about 1 Gyr.
The star formation ceases owing to consumption of the gas.
When the whole system collapses, the number of the clumps
decreases suddenly (see also Fig.\ \ref{anim1}).
This clearly shows that the galaxy is built through
the mergers of sub-clumps, i.e., the hierarchical clustering.
At the same time, the angular momentum of the gas component
decreases well before the gas is transformed into stars.
This provides an evidence that the mergers of the sub-clumps lead to
the angular momentum transfer from the gas component to the dark matter.
We have confirmed that the angular momentum of the dark matter increases,
at the expense of the decrease in that of the baryon component,
which consists of gas and stars.
We can see that the formation of a spherical system supported
by random motions in Model 1 is caused by 
the mergers of the sub-clumps which lead to
the angular momentum loss of the baryon component
and increase the velocity dispersion.

On the other hand, in Model 2, the angular momentum of the baryon component
is almost conserved during the collapse of the system (Fig.\ \ref{h2}).
The star formation begins just when the system has collapsed
and hence the gas density has become high enough. For this reason,
the star formation has a very high rate and a short period.
Model 2 produces a disk-like object,
because the baryon component conserves the angular momentum
unlike Model 1.
Its collapse along the radial direction is halted at the point
where the centrifugal force is balanced by the gravitational force,
whereas the system is able to collapse along the initial rotation
axis without limit.

In the next section we present the properties of the end-products
of the above evolution. In addition, we quantitatively examine
whether each model can reproduce the observed properties of BEGs.

\begin{deluxetable}{cccccccccc}
\footnotesize
\tablecaption{ Model final properties in the $B$ band. \label {tbl-1}} 
\tablehead{
 \colhead{Model} & \colhead{$\log(I_{{\rm e}})$}
 & \colhead{n} 
 & \colhead{$r_{{\rm e}}$} 
 & \colhead{$\log(I_e)$ or $\log(I_0)$\tablenotemark{a}} 
 & \colhead{$r_{{\rm e}}$ or $r_{{\rm D}}$\tablenotemark{b}}
 & \colhead{$M_B$} & \colhead{$V$} 
 & \colhead{$\sigma$}  & \colhead{$V/\sigma$} \\
 \colhead{} & \colhead{($L_{\odot,B}\ {\rm pc }^{-2}$)} & \colhead{}
 & \colhead{(kpc)} & \colhead{($L_{\odot,B}\ {\rm pc }^{-2}$)}
 & \colhead{(kpc)}
 & \colhead{(mag)} & \colhead{(km/s)} & \colhead{(km/s)}
 & \colhead{} 
}
\startdata 
 1 & 1.45 & 4.66 & 8.90 & 1.51 & 8.41 & $-21.37$ & 20 & 167 & 0.12
\nl 
 2 & 2.64 & 1.30 & 3.08 & 3.44 & 1.89 & $-21.38$ & 350 & 217 & 1.61
\enddata
\tablenotetext{a}{For Model 1, $\log(I_e)$ of the $r^{1/4}$ law is listed.
For Model 2, $\log(I_0)$ of the exponential law is listed.}
\tablenotetext{b}{For Model 1, $r_{{\rm e}}$ of the $r^{1/4}$ law
 is listed.
 For Model 2, $r_{{\rm D}}$ of the exponential law is listed.}
\end{deluxetable}

\subsection{Photometric and Kinematic Properties at $z=0$}

 Fig.\ \ref{last} shows final distributions of stellar particles.
These panels show a clear difference between the two configurations.
The end-product of Model 1 has a nearly spherical shape, whereas
that of Model 2 is flat.
These stellar particles contain the information about their
age and metallicity owing to the self-consistent calculation
of the chemical and dynamical evolution.
By means of the population synthesis, we can derive the
photometric properties of the stellar system from this information.
Here, the spectral energy distribution (SED) of each stellar particle
is assumed to be that of a single stellar population (SSP)
that means a coeval and chemically homogeneous assembly of stars.
The SED in a projected region, corresponding to one pixel of CCD,
is obtained by the summation of
the SEDs of the stellar particles whose projections fall
within this region as follows:
\begin{equation}
 F_\lambda = \sum_{i=1}^n m_i f_\lambda(t_i,Z_i),
\end{equation}
where $m_i$, $t_i$, and $Z_i$ are the mass, age, and
metallicity of $i$-th stellar particle and
$f_\lambda(t,Z)$ is the SED of SSP of 1 ${\rm M_\odot}$ with the age,
$t$, and the metallicity, $Z$.
Here we used the data of SSPs of Kodama \& Arimoto (1997).
Since the observational data with which our results should be compared
provide the luminosity distribution projected to a plane,
we first derive the projected distribution of SED
from the three dimensional distribution of stellar particles.
Fig.\ \ref{last} shows the $B$ band images corresponding to the
projected distribution of stellar particles.
Here a 501 $\times$ 501 pixel mesh is chosen
to span the squared region with 50 kpc on a side, and
the flux of each stellar particle is
smoothed using a gaussian filter with the filter scale, $r_{\rm FS}$,
of 1/4 times the softening length of the stellar particle,
i.e., $r_{\rm FS}=0.25 \varepsilon_{\rm s}$.
We discuss the dependence of the results on $r_{\rm FS}$ later.
These images provide quite similar information to the imaging data obtained in
actual observations.
Thus we can obtain various photometric properties from
these images in the same way as in the analysis of observational imaging data.

\begin{figure*}
\epsscale{.9}
\plotone{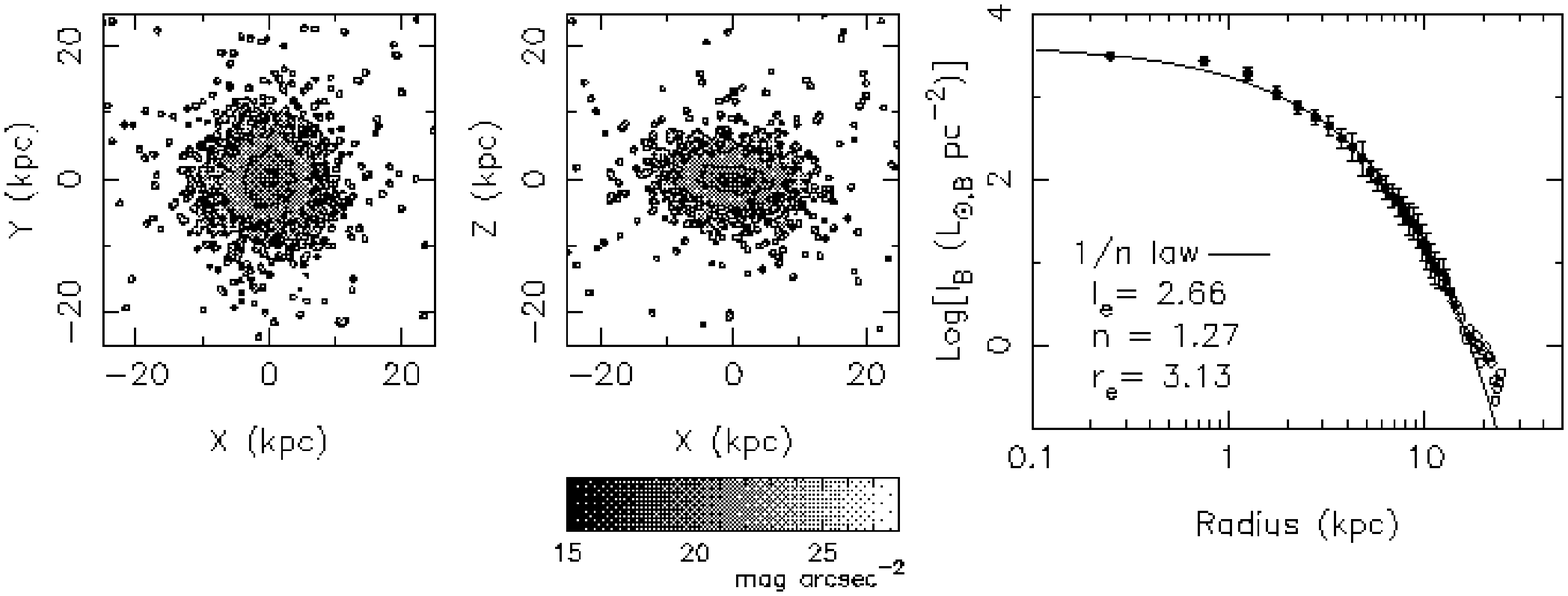}
\figcaption[f7.eps]
{  Final $B$ band images (left and middle) and
the $B$ band surface brightness profile (right) in Model 2.
These are the same as Figs.\ \ref{last} and \ref{sfb},
but the flux of each stellar particle is smoothed with a smaller $r_{\rm FS}$
(for details, see text).
\label{hrs2dppr}}
\end{figure*}

\begin{figure*}
\epsscale{.9}
\plotone{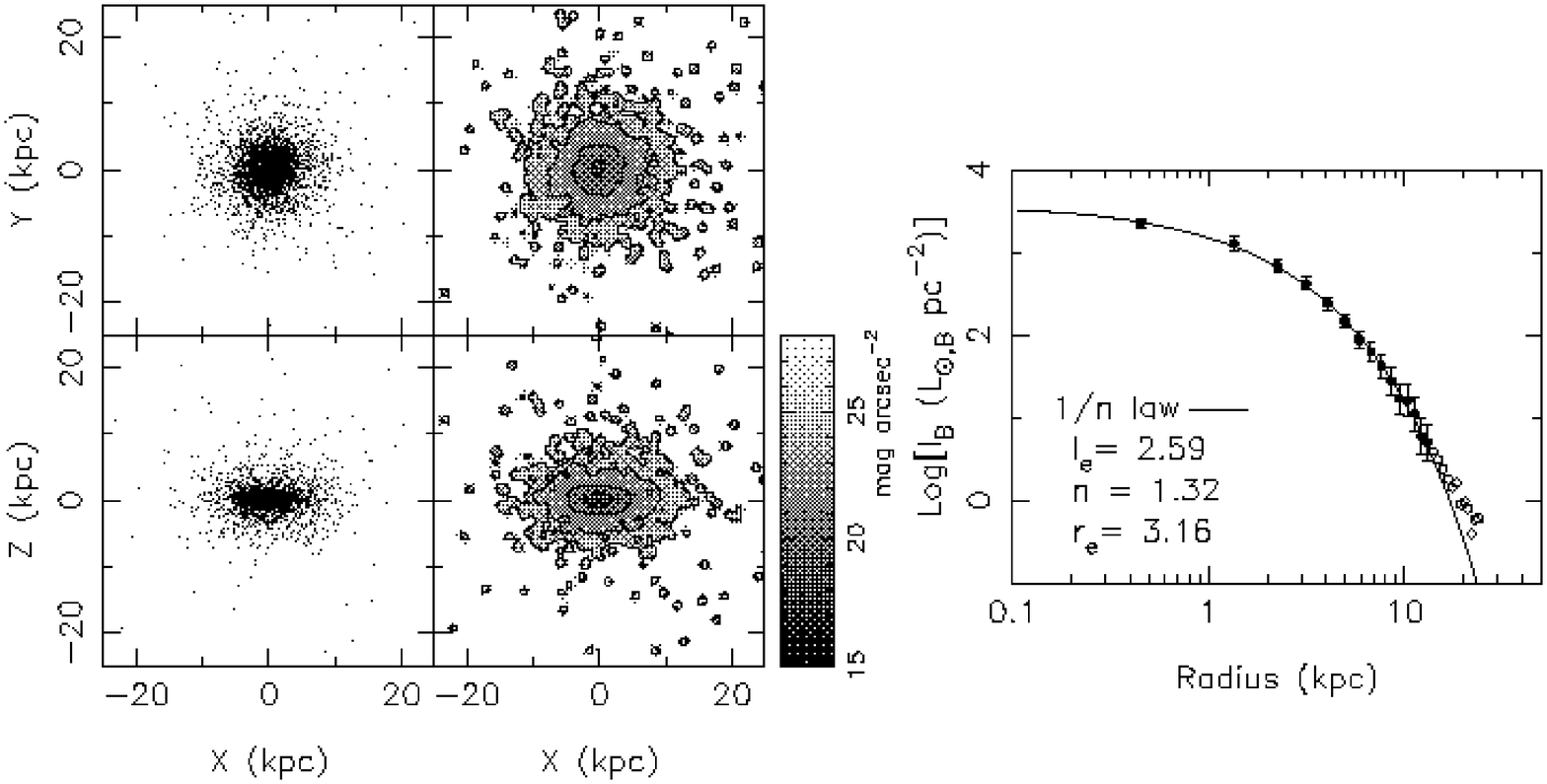}
\figcaption[f8.eps]
{ Final distribution of stellar particles (left), $B$ band image (middle), and
the $B$ band surface brightness profile (right) in Model 2s.
These are the same as Figs.\ \ref{last} and \ref{sfb},
but for Model 2s.
\label{ns2dppr}}
\end{figure*}

 Fig.\ \ref{sfb} shows the azimuthally averaged surface-brightness
profiles obtained from the $x$-$y$
projected $B$ band image shown in Fig.\ \ref{last}.
In getting these profiles, we set the center to the position of a pixel
which has the maximum $V$ band luminosity.
We set the width of each annulus (bin size) which defines the sampling range
to half the softening length.
The error bars correspond to the standard deviation within each
annulus. Since the error is calculated in the flux, like
$\sigma_{f_{B}}^2=\langle f_{B}^2\rangle
-\langle f_{B}\rangle^2$,
the error in the magnitude is defined, using the Taylor expansion, as
\begin{equation}
\sigma_{M_B}=2.5\left(\frac{\sigma_{f_{B}}}{\langle f_{B}\rangle\ln (10)}
-\frac{\sigma_{f_{B}}^2}{2\langle f_{B}\rangle^2 \ln(10)} \right).
\end{equation}
These surface brightness profiles can well be fitted by 
the Sersic ($r^{1/n}$) law,
\begin{equation}
 I(r)=I_e 10^{\{-b_n[(r/r_{{\rm e}})^{1/n}-1]\}},
\label{serpfeq}  
\end{equation}
 where we adopt $b_n=0.868n-0.142$, so that the effective radius,
$r_{\rm e}$,
equals the half light radius in the range $0.5\leq n \leq 16.5$
(Caon et al.\ 1993) and $I_e$ is the surface brightness at $r_{\rm e}$.
This corresponds to the de Vaucouleurs ($r^{1/4}$) law when $n=4$,
whereas the profile with $n=1$ corresponds to the exponential law,
\begin{equation}
 I(r)=I_0 \exp(-r/R_{\rm D}),
\label{exppfeq}  
\end{equation}
where $R_{\rm D}$ is the scale length and
$I_0$ is the surface brightness at the center.
The solid lines in Fig.\ 2 show the best-fits in applying the $r^{1/n}$ law.
The dotted line in the left (right) panel of Fig.\ \ref{sfb} shows the
best-fit in applying the de Vaucouleurs (exponential) law,
i.e., holding $n=4$ ($n=1$) in equation (\ref{serpfeq}).
In fitting, we use the data with $\mu_B < 25.5\ {\rm mag\ arcsec^{-2}}$.
The best-fit parameters are shown in the lower left corner in each
panel as well as in Table \ref{tbl-1}.
 The best-fit $n$ in Model 1 is almost equal to four,
and the de Vaucouleurs law also provides a good fit. 
Therefore, Model 1 succeeds in reproducing the light profile of BEGs,
which is well fitted by the de Vaucouleurs law.
On the other hand, the best-fit $n$ in Model 2 is close to one,
i.e., its profile is well described by the exponential law.
This profile resembles the luminosity distribution of a disk galaxy
rather than that of BEGs. However, the central luminosity
is too large for disk galaxies ($m_{0,B} \sim 18.45$ mag arcsec$^{-2}$).

\begin{figure*}
\epsscale{.9}
\plotone{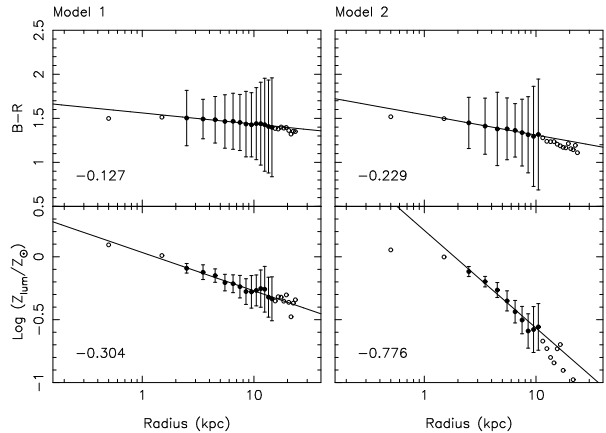}
\figcaption[f9.eps]
{
Color (upper panel) and metallicity (lower panel) gradients
in Model 1 (left) and Model 2 (right).
The solid lines show the best-fit linear relations
for the data plotted as the solid symbols.
The gradients in the best-fit lines are shown in the lower left corner of
each panel. The error bars are shown only for the solid symbols
and correspond to the standard deviation.
\label{grad}}
\end{figure*}

 To see the robustness of these surface-brightness profiles, we examined
the effects of $r_{\rm FS}$ and the number of particles on these profiles.
Here we focus on Model 2, because Model 2 provides a more compact system
and is likely to be affected by the spatial resolution more severely
than Model 1 and the robustness of
the profiles in Model 1 has already been examined in Kawata (1999).
A large $r_{\rm FS}$ is likely to lead to an artificially spread
luminosity distribution and a large half-light radius.
Fig.\ \ref{hrs2dppr} shows the $B$ band images
obtained with $r_{\rm FS}=0.125 \varepsilon_{\rm s}$,
i.e, half of $r_{\rm FS}$ in Figs.\ \ref{last} and \ref{sfb}.
The radial surface-brightness profile obtained with a bin size of
$0.25 \varepsilon_{\rm s}$ are indicated by points.
Although the $B$ band surface-brightness distribution is not smooth,
the best-fitting parameters to the profile agree well with the ones
in Fig.\ \ref{sfb}. This demonstrates clearly that the profile
 in Fig.\ \ref{sfb} is not artificially spread.
In addition, we carried out the same simulation as Model 2
using a smaller number of particles, i.e., 5575 particles for
gas and dark matter respectively. We call this simulation ``Model 2s''.
The softening lengths of gas, stellar, and dark matter particles
are set to $1.82$, $1.82$, and $5.05\ {\rm kpc}$ respectively.
Since Model 2s employs a smaller number of particles and
smaller softening lengths for gas and stars, 
this model has a more favorable condition for the artificial
two-body scattering than Model 2. Fig.\ \ref{ns2dppr} shows 
the luminosity distributions in Model 2s. Here we use
$r_{\rm FS}=0.25 \varepsilon_{\rm s}$.
The artificial two-body scattering is expected to lead
to a more extended surface-brightness profile than that of Model 2.
However, there is little difference
in the images and the profiles between Model 2 and Model 2s.
Therefore we conclude that the numerical resolution in Model 2
is sufficiently fine for the discussion of the luminosity distribution
to be meaningful.

 It is known that a violent relaxation (Lynden-Bell 1967)
in sufficiently deep central potentials reproduces the de Vaucouleurs
law (Hjorth \& Madsen 1991 and references therein).
A deep potential causes a strong scattering
of tightly bound particles passing near the center, and
leads to the outer envelope of the de Vaucouleurs profile.
In addition, the clumpiness can strengthen the violent relaxation so that
particles are thrown to large radii.
van Albada (1982) showed that a cold collapse with
the formation and destruction of substructures is preferred
in producing a de Vaucouleurs profile in dissipationless collapse.
Carlberg, Lake, \& Norman (1986) found
that dissipation, leading to a deep central potential,
could help get a good de Vaucouleurs profile.
Katz \& Gunn (1991) and Katz (1992) presented that
dissipational hierarchical clustering is suitable for bulge formation.
Our results also indicate that a collapse accompanied with the
dissipation and the mergers of the sub-clumps is quite appropriate
for reproducing the de Vaucouleurs profile.

\begin{deluxetable}{ccc}
\footnotesize
\tablecaption{ Color and Metallicity Gradients. \label {tbl-2}} 
\tablehead{
 \colhead{Model} & \colhead{$\Delta(B-R)/\Delta\log(r)$}
 & \colhead{$\Delta\log(Z/Z_\odot)/\Delta\log(r)$} \\
}
\startdata 
 1 & $-0.127$ & $-0.304$
\nl 
 2 & $-0.229$ & $-0.776$
\enddata
\end{deluxetable}

 Fig.\ \ref{grad} shows the color and luminosity weighted
metallicity gradients of the end-products in both the models.
The color gradients are obtained by setting annuli of
various radii in the $B$ and $R$ band images
and subtracting the $R$ band magnitude from the $B$ band magnitude
in each annulus. We made the projected images also for the metallicity,
and obtained their radial profile.
Points indicate the average value in each annulus in the $x$-$y$ projection.
The width of each annulus is set to half the softening length.
The error bars show the standard deviation in each annulus.
The error in the color is written as
$\sigma_{B-R}=\sqrt{\sigma_{M_B}^2+\sigma_{M_R}^2}$.
The profiles of color and metallicity [defined as
$\log (Z/Z_\odot)$] are fitted by linear regression.
In fitting, we excluded the data at radii less than the softening
length and greater than the radius at which the $B$ band surface brightness
is $\mu_B=24.5\ {\rm mag\ arcsec^{-2}}$ (open symbols in Fig.\
\ref{grad}), because the inner region is affected by
the smoothing of gravitational forces whereas the number of particles
within an annulus is too small in the outer region.
The best-fit gradients are shown in the lower left corner
of each panel in Fig.\ \ref{grad} and Table \ref{tbl-2}.
The data obtained from the simulation results
have large errors leading to large errors in the calculated gradients.
However negative gradients are clearly seen in colors
and metallicity, i.e., the color (metallicity) at the center is
redder (higher) than that in the outer region.
 From the observational studies, it is known that typical BEGs have
$\Delta(B-R)/\Delta\log(r)=-0.09\pm0.02$ (Peletier et al.\ 1990)
and $\Delta\log(Z/Z_\odot)/\Delta\log(r)=-0.30\pm0.12$
(Kobayashi \& Arimoto 2000 and references therein).
In Model 1 the gradients of both the color and the metallicity
are in fairly good agreement with the observed gradients of BEGs.
On the contrary, these gradients are too steep in Model 2.
Model 1 is thus favored over Model 2 in reproducing 
the color and metallicity gradients as well as the surface-brightness
profiles.

\begin{center}
\epsscale{.4}
\plotone{f10.eps}
\figcaption[f10.eps]
{ The half-light radius (upper panel) and $V/\sigma$ (lower panel) against
the absolute magnitude of the galaxy.
The results of Model 1 and Model 2 are plotted as
a circle and a square respectively. Crosses indicate the data
for the ellipticals in the Virgo and Coma clusters
(Faber et al.\ 1989; Davies et al.\ 1983).
\label{mbrev}}
\end{center} 

Carlberg (1984) showed that the dissipational collapse
of a proto-galaxy with a uniform density leads to
a steep metallicity gradient of
$\Delta\log(Z/Z_\odot)/\Delta\log(r)\sim-0.5$,
using the numerical simulation. It has already been a serious problem
that a metallicity gradient expected in the uniform collapse
becomes steeper than the observed one
(e.g., Davies, Sadler, \& Peletier 1993; Kobayashi \& Arimoto 2000).
The negative gradients in both the color and the metallicity
mean that the star formation persists
over longer time in the inner region than in the outer region,
due to dissipative infall of gas. Since the residual gas is polluted
by the past star formation, metal-rich stars are formed in
the central region (Kawata 1999).
In Model 1 the star formation which occurs
in the sub-clumps before the whole system collapses decreases
the amounts of the residual gas infalling dissipatively, leading
to an appropriate shallower gradients.
In addtion, the stellar mergers are expected to weaken an existing gradient
(White 1980). Davies et al.\ (1993)
supposed that BEGs have to experience some stellar merger events.
The scattering induced by the clustering of the stellar sub-clumps
(Fig. \ref{anim1}) in Model 1 is also likely to play
a quite important role of diluting
the metallicity gradient which leads to the color gradient.
The shallower gradients of Model 1 seem to be caused by the
combination of these two effects.
The results of our realistic numerical simulations suggest that
the formation and clustering of sub-clumps
are preferred in producing the observed color and metallicity gradients.

 In Fig.\ \ref{mbrev} we compare the sizes and the kinematical properties
of the final stellar systems with those of the observed BEGs.
The upper and lower panels respectively plot
the half-light radii, $r_{{\rm e},B}$,
and the ratios of the rotation velocity to the average velocity dispersion
within $0.5 r_{{\rm e},B}$, $V/\sigma$,
against the $B$ band absolute magnitude.
The data of elliptical galaxies in
the Virgo and Coma clusters are plotted as crosses,
where we used the total magnitudes and the half-light radii
of Faber et al.\ (1989), and $V/\sigma$ of Davies et al.\
(1983), assuming distance moduli of
31.0 for the Virgo (Graham et al.\ 1999) and
34.7 for the Coma (Bower et al.\ 1992).
There is no difference between Model 1 and Model 2
in the total magnitudes of the final objects.
However, the half-light radius in Model 1
is much larger than that in Model 2 and is in good agreement with
that of the observed elliptical galaxies with the same luminosity.
As to the kinematic properties,
Model 1 succeeds in reproducing a low $V/\sigma$ of BEGs very well.
Owing to the loss of the angular momentum (Fig.\ \ref{h1}),
the end-product in Model 1 is a system
supported by the random motions and hence rotating slowly.
Furthermore, the small-scale initial density fluctuations
allow star formation in the sub-clumps
before the whole system begins to collapse and
the collisionless property of the stellar component prevents
the dissipative contraction of the system,
leading to an appropriate half-light radius in Model 1.
On the other hand, in Model 2
the end-product is a nearly rotation-supported system,
owing to the angular momentum conservation (Fig.\ \ref{h2}).
In addition, the initial-small angular momentum
requires a large contraction along the radial direction
for producing enough centrifugal force, leading to 
a small half-light radius.
These results demonstrate that, while
Model 2 is clearly unacceptable, Model 1
is pretty appropriate for producing BEGs.

\section{Conclusion}

 We showed that the luminosity profile, color and metallicity gradients,
half-light radius, and $V/\sigma$ of 
the end-product formed by clustering of sub-clumps
are in very good agreement with those of the observed BEGs
quantitatively. Since the CDM cosmology naturally predicts
the existence of massive low-spin seed galaxies including
small-scale density fluctuations,
the CDM scenario is quite suitable for the formation of BEGs
from a massive low-spin seed galaxy, in which the clustering
of sub-clumps plays an essential role.

 On the other hand, in a massive low-spin seed galaxy including
no initial small-scale density fluctuation, the small initial spin
leads to a compact disk-like object dominated by the rotation
owing to the angular momentum conservation.
The disk has an exponential light profile,
which is obviously inconsistent with the light profiles
of the observed BEGs. Furthermore, the central surface brightness
is too high to be in agreement with that of disk galaxies.
No object like this has been observed in our universe
despite its high surface brightness.
This result suggest that a massive low-spin seed galaxy
is not allowed to have a uniform density.

\acknowledgments
DK would like to thank Masafumi Noguchi for invaluable discussion
and carefully reading the manuscript.
DK is grateful to Nobuo Arimoto and Tadayuki Kodama
for kindly providing the tables of their SSPs data.
DK also acknowledges
the Yukawa Institute Computer Facility,
the Institute of Space and Astoronautical Science,
and the Astronomical Data Analysis Center of the National Astronomical
Observatory, Japan
where the numerical computations for this paper were performed.
This work is supported in part by
the Japan Atomic Energy Research Institute.


\begin{references}
\reference{} Barnes, J., \& Efstathiou, G.\ 1987, \apj, 319, 575
\reference{} Barnes, J.E., \& Hut, P.\ 1986, Nature, 324, 446
\reference{} Bertschinger, E.\ 1995, preprint (astro-ph/9506070)
\reference{} Bower, R.G., Lucey, J.R., \& Ellis, R.S.\ 1992,
 \mnras, 254, 601
\reference{} Caon, N., Capaccioli, M.,
 \& D'Onofrio, M.\ 1993, \mnras, 265, 1013
\reference{} Carlberg, R.G.\ 1984, \apj, 286, 403
\reference{} Carlberg, R.G., Lake, G., \& Norman, C.A.\ 1986, \apj,
 300, L1
\reference{} Davies, R.L., Efstathiou, G., Fall, M., Illingworth, G.,
 \& Shechter, P.L.\ 1983, \apj, 266, 41
\reference{} Davies, R.L., Sadler, E.M., \& Peletier, R.F.\ 1993,
 \mnras, 262, 650
\reference{} Eisenstein, D.J., \& Loeb, A.\ 1995, \apj, 439, 520
\reference{} Faber, S.M., Wegner, G., Burstein, D., Davies, R.L.,
 Dressler, A., Lynden-Bell, D., \& Terlevich, R.J.\ 1989, \apjs, 69, 763
\reference{} Gingold, R.A., \& Monaghan, J.J.\ 1977, \mnras, 181, 375
\reference{} Graham, J.A., Ferrarese, L., Freedman, W.L., Kennicutt, R.C.Jr.,
 Mould, J.R., Saha, A., Stetson, P.B., Madore, B.F.,
 Bresolin, F., Ford, H.C., Gibson, B.K., Han, M., Hoessel, J.G.,
 Huchra, J., Hughes, S.M. Illingworth, G.D., Kelson, D.D., Macri, L.,
 Phelps, R., Sakai, S., Silbermann, N.A., \& Turner, A.\ 1999, \apj,
 516, 626
\reference{} Heavens, A., \& Peacock, J.\ 1988, \mnras, 232, 339
\reference{} Hernquist, L., \& Katz, N.\ 1989, \apjs, 70, 419
\reference{} Hjorth, J., \& Madsen, J.\ 1991, \mnras, 253, 703
\reference{} K\"aellander, D., \& Hultman, J.\ 1998, A\&A, 333, 399
\reference{} Katz, N.\ 1992, \apj, 391, 502
\reference{} Katz, N., \& Gunn, J.E.\ 1991, \apj, 377, 365
\reference{} Katz, N., Weinberg, D.H., \& Hernquist, L.\ 1996, \apjs, 105, 19
\reference{} Kawata, D.\ 1999, \pasj, 51, 931
\reference{} Kay, S.T., Pearce, F.R., Jenkins, A., Frenk, C.S.,
 White, S.D.M., Thomas, P.A., \& Couchman, H.M.P.\ 2000, \mnras, 316, 374
\reference{} Kobayashi, C., \& Arimoto, N.\ 2000, \apj, 527, 573
\reference{} Koda, J., Sofue, Y., \& Wada, K.\ 2000a, \apj, 531, L17
\reference{} Koda, J., Sofue, Y., \& Wada, K.\ 2000b, \apj, 532, 241
\reference{} Kodama, T., \& Arimoto, N.\ 1997, A\&A, 320, 41
\reference{} Lucy, L.B. 1977, \aj, 82, 1013
\reference{} Lynden-Bell, D.\ 1967, \mnras, 136, 101
\reference{} Padmanabhan, T. 1993, Structure formation in the universe
 (Cambridge: Cambridge Univ.\ Press)
\reference{} Peebles, P.J.E.\ 1971, A\&A, 11, 377
\reference{} Peletier, R.F., Davies, R.L., Illingworth, G.D.,
 Davis, L.E., \& Cawson, M.\ 1990, \apj, 100, 1091
\reference{} Steinmetz, M., \& Bartelmann, M.\ 1995, \mnras, 272, 570
\reference{} Steinmetz, M., \& M\"uller, E.\ 1994, A\&A, 281, L97
\reference{} Steinmetz, M., \& M\"uller, E.\ 1995, \mnras, 276, 549
\reference{} Theis, Ch., Burkert, A., \& Hensler, G.\ 1992, A\&A, 265, 465
\reference{} van Albada, T.S.\ 1982, \mnras, 201, 939
\reference{} Warren, M., Quinn, P.J., Salmon, J.K.,
 \& Zurek, W.H.\ 1992, \apj, 399, 405
\reference{} White, S.D.M.\ 1980, \mnras, 191, 1
\reference{} White, S.D.M., \& Frenk, C.S.\ 1991, \apj, 379, 52
\end{references}
\end{document}